\documentclass{iaus}
\usepackage{graphicx}

\title[magnetic fields - torsional waves] 
{Correlations of magnetic features and the torsional pattern}

\author[Judit Murak\"ozy \& Andr\'as Ludm\'any]   
{Judit Murak\"ozy \and Andr\'as Ludm\'any}

\affiliation{Heliophysical Observatory, H-4010 Debrecen P.O.Box 30, Hungary 
}
\pubyear{2010}
\volume{273}  
\pagerange{119--126}
\jname{Physics of Sun and Star Spots}
\editors{A.C. Editor, B.D. Editor \& C.E. Editor, eds.}
\begin{document}

\maketitle

\begin{abstract}

The striking similarity between the cyclic equatorward migration of the torsional oscillation belts and the sunspot activity latitudes inspired several attempts to seek an explanation of the torsional phenomenon in terms of interactions between flux ropes and plasma motions. The aim of the present work is to examine the spatial and temporal coincidences of the torsional waves and the emergence of sunspot magnetic fields. The locations of the shearing latitudes have been compared with the distributions of certain sunspot parameters by using sunspot data of more than two cycles. The bulges of the sunspot number and area distributions tend to be loacted within the 'forward' belts close to their poleward shearing borders. A possible geometry of the magnetic and velocity field interacion is proposed.

\keywords{torsional wave, sunspots}
\end{abstract}

\firstsection 

\section{Introduction}

The torsional oscillation was firstly reported by Howard and LaBonte {1980} as a travelling wave pattern superposed on the differential rotation. At the beginning the similarity of the equatorward migration of these belts and the butterfly diagram was very remarkable and inspiring, Labonte and Howard (1982) already provided a comparison of the belts and the latitudinal distribution of the emerging magnetic field by using magnetograms. Snodgrass (1987a, 1987b), Wilson (1987), Snodgrass and Wilson (1987) presented arguments that the torsional pattern is related to large convection cells. Subsurface measurements were reported by Kosovichev and Schou (1997) and Schou et al. (1998), Howe et al., (2000a) published results of GONG and MDI subsurface measurements and reported torsional pattern down to about 0.92 R$_{\odot}$, this was confirmed by Komm et al. (2001) and Antia and Basu (2001). The torsional oscillation can be regarded to be a persistent feature extending down to about 0.92 $R_{\odot}$.

The early models followed global approach, the works of Yoshimura (1981) and Sch\"ussler (1981) were based on the Lorentz force. Later models followed local treatment. K\"uker et al. (1996) considered magnetic quenching of Reynolds stresses modifying the differential rotation profile.  In the model of Petrovay and Forg\'acs-Dajka (2002) the sunspots modify the turbulent viscosity in the convective zone which leads to the modulation of the differential rotation. 

The present paper aims at finding any spatial correlation between the torsional wave and any sunspot feature. Earlier works also indicated some spatial connections but only based on magnetograms. LaBonte and Howard (1982) averaged the latitudinal magnetic activity distribution for a longer time, whereas Zhao and Kosovichev (2004) only indicated the location of the activity belt with no distribution information. By using sunspot data, one can study the role of the most intensive magnetic fluxes hopefully leading to an answer for the question: which properties of the magnetic flux ropes could be able to modify the ambient flow resulting in the observed zonal velocity pattern.

\section{Observational material and method of analysing}

The aim is to follow the temporal variation of the latitudinal distribution of sunspot parameters in comparison with the migration of the torsional waves. Four parameters were considered: the number and the total area of sunspots as well as the mean number of spots in groups and the mean area of groups. These latter two parameters can characterise the complexity of the sunspot groups to check the assumption that larger clusters of flux ropes can modify more efficiently the ambient velocity fields. 

\begin{figure}[ht]
\begin{center}
 \includegraphics[width=3.5in]{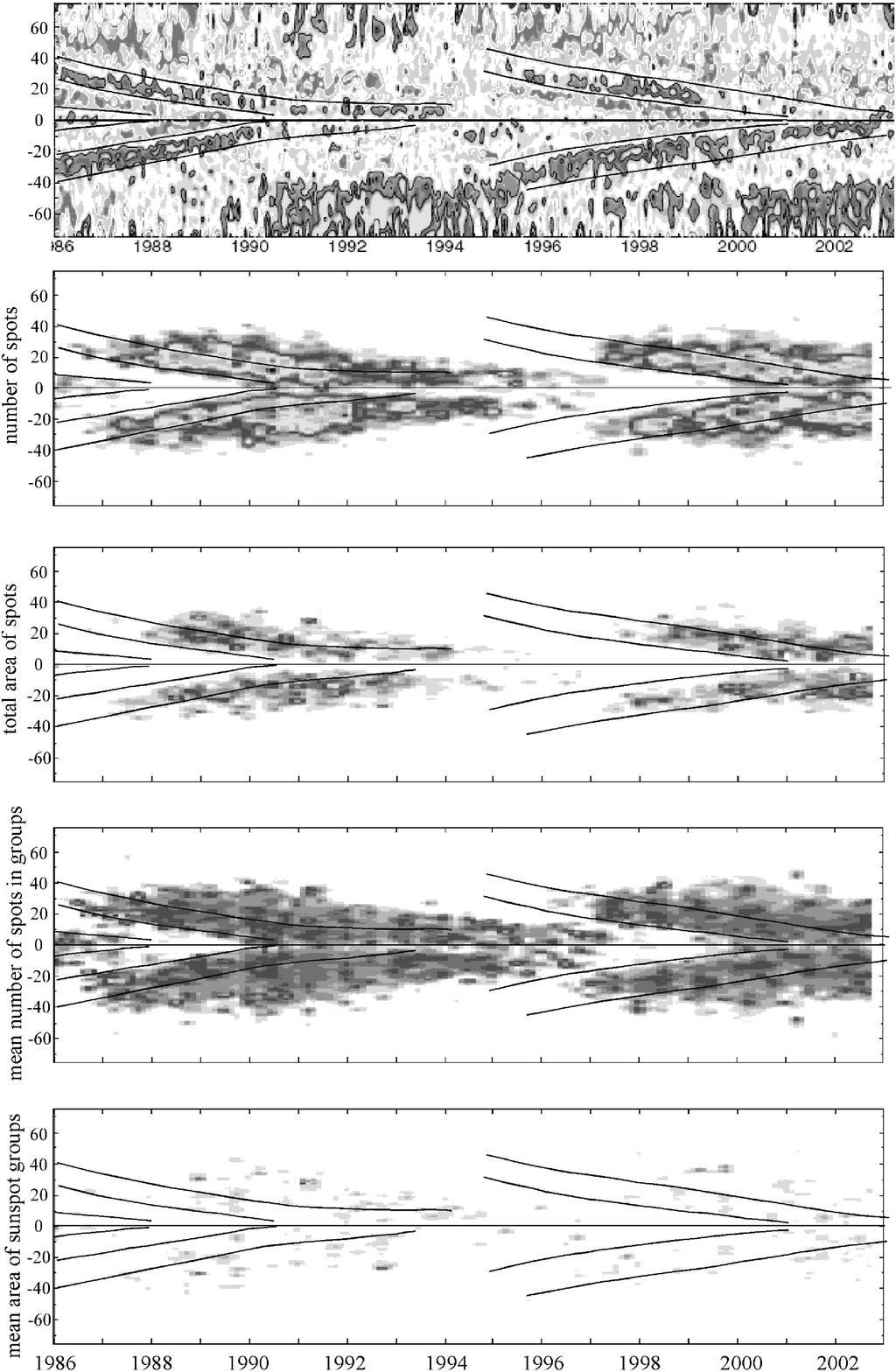} 
 \caption{The upper panel shows the azimuthal velocity data of Ulrich and Boyden (2005) in black-and-white reproduction along with additionally inserted borderlines of the prograde belts. These borderlines are also inserted into the rest of the panels. The second and third panels show the distributions of the number and the area of spot groups. The fourth and fifth panels show the mean number of spots in the groups and the mean area of groups.}
   \label{mw-db}
\end{center}
\end{figure}

The sunspot data were taken from the most detailed sunspot database, the Debrecen Photoheliographic Data (DPD, Gy\H ori et al, 2010). The present study analyses the years 1986-2000. The latitudinal distributions were determined in such a way that latitudinal stripes of one degree were considered, and all applied sunspot parameters were summed up for all stripes on a three-monthly basis. 

To compare the obtained distributions with the torsional pattern one has to determine the latitudinal location of the torsional wave i.e. the latitudes of fast/slow belts and the shearing zones. The torsional data were taken from the paper of Ulrich and Boyden (2005), Figure 1. shows these data for the period 1986-2002. Since fractal-like distributions cannot be compared directly we inserted curves on the shear zones indicating the prograde belts and the same curves were inserted in the four panels of sunspot distributions. The number and the total area of sunspot groups are plotted in the 2-3 panels, the areas are taken into account at the time of the largest extension of the groups. The mean number of spots within the groups as well as the mean area of spot groups are plotted in panels 4 and 5. These last two features characterize the significance of the groups, their size and complexity. They are used to check the assumption that larger and more complex flux rope systems might influence the ambient flows more efficiently and this could be the reason of the torsional pattern (Petrovay, 2002).

\section{Possible scenario of a flow system}

The panels do not support the assumption that the torsional belts are caused by the sunspots themselves because the onset of the belts precedes the  appearance of the first spot in the new cycles and the complexity of spot groups does not seem to play any role, at least in the present representation of spot group complexity. The following properties of the distributions are conspicuous:

1. The line of weight of the area occupied by sunspots is located on the borderline between the prograde and the poleward retrograde belts.

2. The equatorward borderline of the prograde belt is a definite borderline of the sunspot occurrence (see the second panel of Fig.1).

\begin{figure}[ht]
 \begin{center}
 \includegraphics[width=3in]{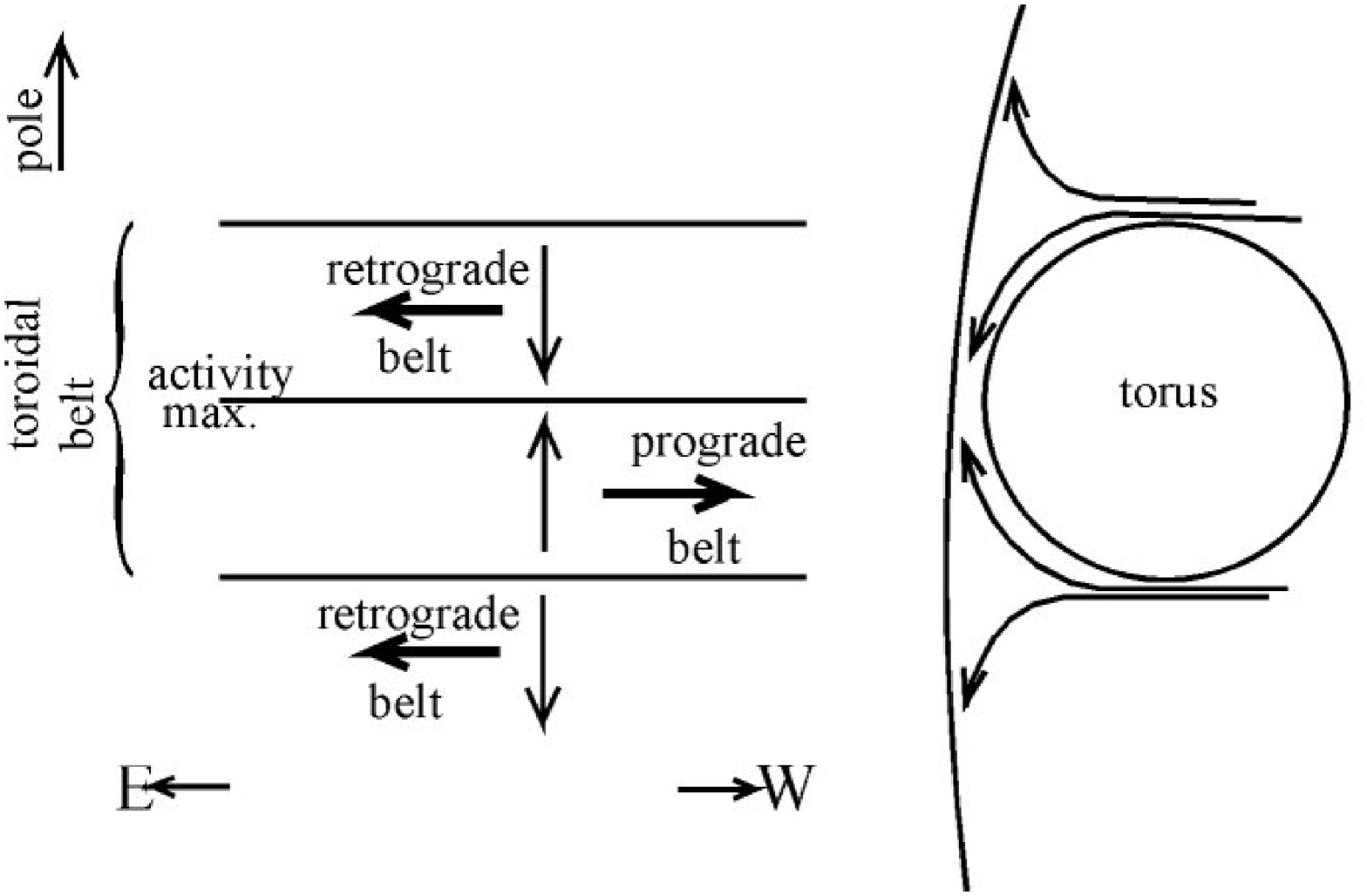}
 \caption{Schematic representation of the proposed mechanism in the northern hemisphere. Left: view from outside, right: view in the meridional plane from west.}
   \label{band}
 \end{center}
\end{figure}

The second feature may be a signature of certain internal streams, it seems to allow the following interpretation. The cluster of toroidal magnetic field ropes may be present in the convective zone up to a certain height and may act as an obstacle for the emerging material. Out of this region (poleward and equatorward from the toroidal belt) the uprising streams may be stronger than above the torus and they turn around the obstacle, the torus. Above the torus these turning around streams flow equatorward in the poleward half of the toroidal belt and they flow poleward in the equatorward half. The former flow is then oriented eastward and the latter one westward due to the Coriolis force, these produce the retrograde/prograde belts, see Fig.\,\ref{band}. The inward motion at active regions has been observed by Zhao and Kosovichev (2004). This mechanism avoids the problem of the apparent violation of causality arising in models based on sunspots. In this model the source region of the sunspots, i.e. the toroidal field is the cause of the torsional pattern which is present prior to the emergence of the first sunspot in a cycle. This restricts the onset of the belts to middle latitudes.

\begin{figure}[ht]
 \begin{center}
 \includegraphics[width=2in]{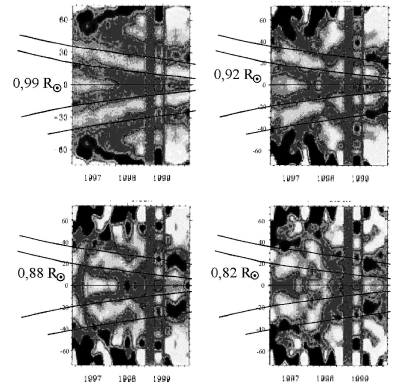}
 \caption{Internal torsional belts between 1996-1999 reported by Howe et al. (2000) in black-and-white reproduction. The borderlines of Fig.\,\ref{mw-db}Fig.1 are also inserted.}
 \label{internal}
\end{center}
\end{figure}

If this scenario is correct it may have implications for the depth of the toroidal field rope clusters. Howe et al. (2000) reported internal torsional patterns and found that the belts get disintegrated at a depth of about 0.92$_{\odot}$. This may mean that this is the depth of the outermost toroidal field rope clusters above which the torsional belts can be formed.

\begin{acknowledgements}

The research leading to these results has received funding from the European Community's Seventh Framework Programme (FP7/2007-2013) under grant agreement No. 218816. 

\end{acknowledgements}

\end{document}